\documentclass[letterpaper,twocolumn,amsmath,amssymb,prl]{revtex4}

\usepackage{graphics}
\usepackage{graphicx}
\usepackage{dcolumn}
\usepackage{bm}
\usepackage{amsmath}
\usepackage{rotating}
\usepackage[bookmarks = true, pdfstartview = FitH, colorlinks = true, urlcolor = blue,hyperfootnotes=false]{hyperref}

\newcommand{\up}{\mathord{\uparrow}}
\newcommand{\down}{\mathord{\downarrow}}
\newcommand{\bra}[1]{\langle #1|}
\newcommand{\ket}[1]{|#1\rangle}

\renewcommand{\v}[1]{\mathbf{#1}}

\begin{document}
\title{Sideband cooling while preserving coherences in the nuclear spin state in Group-II-like atoms}
\author{Iris Reichenbach}
\email{irappert@unm.edu}
\affiliation{{Department of Physics and Astronomy, University of New Mexico, Albuquerque, New Mexico 87131, USA}}
\author{Ivan H. Deutsch}
\affiliation{{Department of Physics and Astronomy, University of New Mexico, Albuquerque, New Mexico 87131, USA}}

\begin{abstract}
We propose a method for laser cooling group-II-like atoms without changing the quantum state of their nuclear spins, thus preserving coherences that are usually destroyed by optical pumping in the cooling process. As group-II-like atoms have a $^1S_0$ closed-shell ground state, nuclear spin and electronic angular momentum are decoupled, allowing for their independent manipulation.  The hyperfine interaction that couples these degrees of freedom in excited states can be suppressed through the application of external magnetic fields.  Our protocol employs resolved-sideband cooling on the forbidden clock transition, $^1S_0 \rightarrow {}^3P_0$, with quenching via coupling to the rapidly decaying $^1P_1$ state, deep in the Paschen-Back regime.  This makes it possible to laser-cool neutral atomic qubits without destroying the quantum information stored in their nuclear spins as shown in two examples, $^{171}$Yb and $^{87}$Sr.
\end{abstract}

\maketitle
The ability to coherently control atoms has allowed for major advances in precision measurement for tests of fundamental physics \cite{Insight} and new proposed applications such as quantum simulators and universal quantum computers \cite{Jacksch}.  Though traditionally alkali-metal (group-I) atoms are used in many of these applications given our ability to laser cool and trap them with relative ease, alkaline-earth-metal (group-II) atoms are also being considered given long-lived metastable states, accessible lasers for cooling and trapping, and a rich (but tractable) internal structure.  One distinguishing feature of group-II vs.\ group-I elements is that the ground state is a closed-shell $^1S_0$ with zero electron angular momentum, and thus no hyperfine coupling to the nuclear spin.  This will allow us to independently manipulate nuclear and electronic degrees of freedom in ways that are not easily accomplished with group-I atoms. 

From the perspective of storing and manipulating quantum information, nuclear spins in atoms are attractive given their long coherence times and the mature techniques of NMR \cite{ChuangRMP}.  Recently, we have studied a protocol in which nuclear spins can mediate quantum logic solely through  quantum statistics via an ``exchange blockade" \cite{Hayes06}.  This allows us to isolate nuclear spins from noisy perturbing forces while simultaneously providing a mechanism by which strong qubit-qubit interaction can be implemented via cold collisions.  As is typical in atomic quantum logic protocols, heating of atomic motion degrades performance.  In atoms with group-I-like electronic structures, laser cooling cannot be used to re-initialize atomic vibration in the course of quantum evolution because this is accompanied by optical pumping that erases the qubit stored in the internal degrees of freedom.  One must therefore resort to sympathetic cooling with another species \cite{Sympathetic}.  In this Letter we propose and analyze a new protocol in which one can continuously refrigerate atomic motion while simultaneously maintaining quantum coherence stored in nuclear spins.  Our proposal is based on resolved-sideband laser cooling \cite{diedrich89} of group-II-like atoms, tightly trapped in the Lamb-Dicke regime, as in optical lattices currently under consideration for next-generation atomic clocks based on optical frequency standards \cite{Takamoto05}.  We will explain our cooling scheme in detail using two example species,  ${}^{171}$Yb  and ${}^{87}$Sr,  both of which are actively studied in the laboratory for application to optical clocks \cite{Ludlow06, Yb_clock}.  The $I=1/2$ nucleus of $^{171}$Yb is a natural qubit for storing quantum information.  In the case of $^{87}$Sr, whose nucleus has spin $I=9/2$, we will choose two sublevels in the 10-dimensional manifold to encode a qubit.  Exquisite optical control of the nuclear spin of $^{87}$Sr has recently been demonstrated by Boyd {\it et al.} \cite{Boyd}.

In order to preserve the quantum state of a qubit encoded in a nuclear spin while laser cooling, the spin coherences must be transferred in both excitation and spontaneous decay.  Optical fields interacting with atoms via the electric dipole couple directly to the electrons, affecting nuclear spin states only indirectly via the hyperfine interaction.  A key requirement of our protocol is, therefore, to excite states with negligible hyperfine coupling, and/or decouple electrons from nuclear spin through the application of a sufficiently strong magnetic field that the Zeeman effect on the electrons dominates over the hyperfine interaction (Paschen-Back regime \cite{Foot}).  Moreover, we seek to recool atoms in traps to near the ground vibrational state which requires resolvable motional sidebands.

The dual requirement of very narrow linewidths and small hyperfine coupling leads us to consider cooling on the weakly allowed intercombination  ``clock" transition ${}^1S_0 \rightarrow {}^3P_0$ in a ``magic wavelength" optical lattice \cite{Takamoto05}.  Cooling to the ground vibrational state can proceed by coherent excitation of a $\pi$ pulse on the first red sideband, $\ket{^1S_0,n}\rightarrow\ket{^3P_0,n-1}$, where $n$ is the vibrational quantum number, followed by recycling to the ground state with Lamb-Dicke suppression of recoil.   Relatively large vibrational frequencies of $\omega_v/2 \pi=90 \,(260)$ kHz have already been realized for Yb (Sr)  in such an optical lattice in 1D \cite{Barber06, Lemonde06}.  In principle, given these tight confinements and the tiny linewidth $\gamma$ of the clock transition (few millihertz), the motional sidebands are very well resolved, and extremely cold temperatures can be reached with a steady state mean vibrational excitation on the order of  $\langle n\rangle \approx \gamma^2/(2 \omega_v)^2=10^{-15}\,(10^{-18})$ \cite{diedrich89}. Of course, the true achievable temperature will depend strongly on the suppression of the different heating mechanisms. 

Our concern is to carry out this cooling while maintaining nuclear spin coherence.  To good approximation, the $^3P_0$ state has total electron angular momentum $J=0$, and under this condition, there is no interaction between electrons and nuclear spin, as in the ground state, $\ket{^1S_0}$. However, the clock transition is $J=0 \rightarrow J=0$, and laser excitation is only allowed because in the excited state, the hyperfine interaction leads to a small admixture of the higher-lying $P$ states \cite{Porsev04, Boyd}.  The nuclear spin projection $m_I$ is thus not an exact quantum number for the magnetic sublevels in $\ket{^3P_0}$; a very small admixture of electronic angular momentum renders $m_F$ a good quantum number.  This leads to two effects which we must address in the context of transferring the nuclear spin state. While the clock transition is only allowed due to the hyperfine interaction, it is still the case that the sublevels of  $\ket{^3P_0}$ are almost pure $m_I$, and a $\pi$-polarized pulse will preserve nuclear spin projections in excitation on the clock transition \cite{Boyd}.   More importantly, even a small admixture of electron angular momentum can strongly affect the magnetic moment, so that the $g$-factor of the excited state differs from that of the ground state \cite{Boyd}.  This implies that in a bias magnetic field, the $\pi$-transitions $\ket{^1S_0,n}\otimes\ket{m_I}\rightarrow\ket{^3P_0,n-1}\otimes\ket{m_F=m_I}$ will have different resonant frequencies for different values of $m_I$.  The protocol for the transfer of nuclear spin coherence from the ground state to the clock state will thus depend on the details of the experimental conditions, i.e., the relative size of the differential Zeeman splitting in the given bias field when compared to the vibrational spacing in the trap.  Under any operating condition, these coherences can be transfered due to the extremely narrow linewidth of the clock transition, be it through sequential addressing of each sublevel in a series of narrow band pulses \cite{Boyd}, or in a short pulse that does not resolve the differential Zeeman shift but does resolve the sidebands.  Finally, in this step, one must also take into account that the differential $g$-factor leads to differential relative phases between the sublevels; this is a unitary transformation which can be reversed. 

\begin{figure}
\begin{center}
\includegraphics[width=7cm]{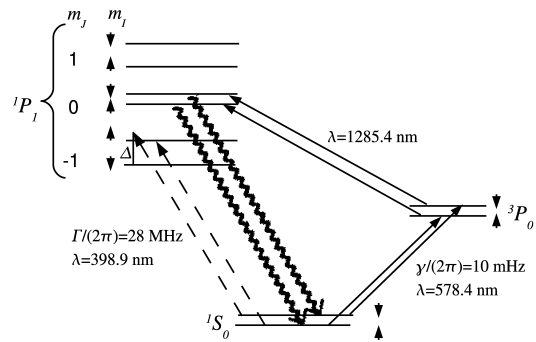}
\caption{Schematic cooling and readout level-diagram for $^{171}$Yb (analogous for $^{87}$Sr).   Sideband cooling in a trap occurs via excitation on the $^1S_0 \rightarrow {}^3P_0$ clock transition (vibrational levels not shown), quenched by coupling to the $^1P_1$ state.  An external magnetic field splits the $^1S_0$ and $^3P_0$ nuclear sublevels (spin-up and spin-down shown) and ensures preservation of the nuclear spin during resonant excitation and repumping.  The dashed arrows demonstrate a possible readout scheme by resonance fluorescence.}
\label{fig:overview}
\end{center}
\end{figure}

Once the atom is transferred to the clock state $^3P_0$, the cooling cycle must be completed through a spontaneous event that returns the atom to its ground electronic state $^1S_0$, with Lamb-Dicke suppression of recoil.  Of importance in our protocol is that this occurs without decoherence of the nuclear spin state.  The key requirement for spontaneous transfer of coherences is that the decay paths of the different sublevels are {\em indistinguishable}, i.e. the decay channels cannot be distinguished by their polarization or frequency.  Under circumstances in which the nuclear spin and electronic degrees of freedom of an excited state are decoupled (no hyperfine interaction), in a product state with zero projection of electron angular momentum, $\ket{e} \approx \ket{I,m'_I} \otimes \ket{J',m'_J=0}$, all decay channels to the ground state are indistinguishable.  To see this note that the selection rules dictate that the electric dipole matrix element  satisfies, $\bra{e}d_q \ket{g}=\bra{J',m'_J=0}d_0\ket{J=0,m_J=0}\delta_{q,0} \delta_{m_I,m'_I}$.  Only $\pi$-transitions are allowed and all decay channels have the same polarization without change of nuclear spin projection.  Moreover, in the $m'_J=0$ manifold, the $g$-factor is solely nuclear, $g_I$, equivalent to that in the ground state.  Thus, the Zeeman splitting in excited and ground states will be equal and decay channels will not be distinguished by frequency.  Deviations of the excited state from a nuclear-electron product state (due to residual hyperfine interaction) will lead to decoherence when there are nuclear spin changing decays (polarization distinguishable) and/or when the different channels are frequency resolvable (i.e. have a frequency difference that is not negligible compared to the decay linewidth).

In the case of the clock state $^3P_0$, spontaneous decay is completely due to the admixture of higher $P$ states and the decay channels with different polarizations occur with a probability proportional to the respective Clebsch Gordan coefficients. Therefore, the nuclear spin is not preserved during spontaneous decay. Moreover, the long lifetime makes the cooling cycle too long for practical purposes.  We thus consider quenching the clock state by pumping to the short lived  ${}^1P_1$ state (see Fig.\ \ref{fig:overview}) \cite{Sterr2004,Curtis2003}.  This level decays with very high probability to the ground state $^1S_0$ and has a very broad linewidth, $\Gamma/(2 \pi) >10$ MHz.  Direct excitation ${}^3P_0 \rightarrow {}^1P_1$, is weakly magnetic-dipole-allowed, and has been considered in the context of electromagnetically induced transparency \cite{Santra05}.  Alternatively, we can quench the clock state via a two-photon $^3P_0 \rightarrow {}^3S_1  \rightarrow{} ^1P_1(m_J=0)$ transition, off resonance from the intermediate state. This has the advantage that, though one leg is an intercombination line, all transitions are electric-dipole allowed, leading to larger depopulation rates for the same intensity.

\begin{figure}
\begin{center}
\includegraphics[width=7.5cm]{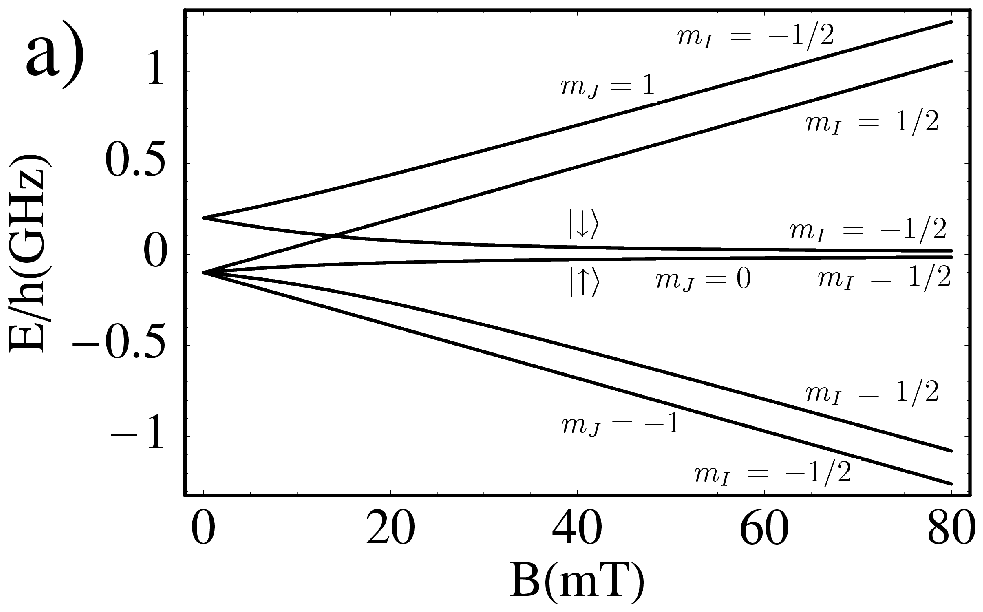}
\includegraphics[width=7.5cm]{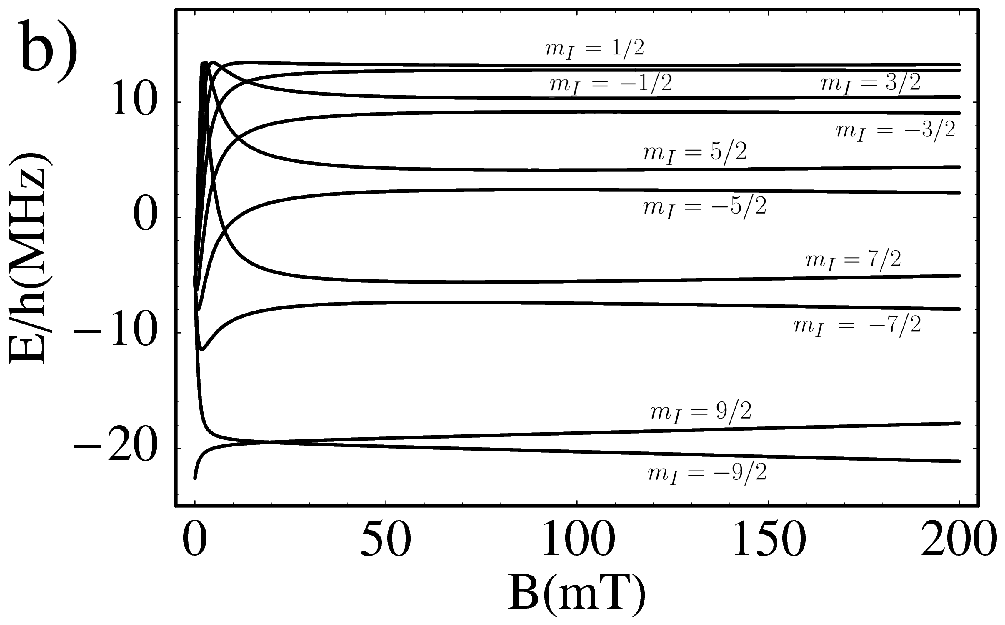}
\caption{Zeeman diagram of the $^1P_1$ manifold. (a) In $^{171}$Yb the Breit-Rabi formula applies. The hyperfine sublevels decouple to product states of electron and nuclear spin (Paschen-Back) with modified linear Zeeman shift. The qubit is encoded the states $\ket{\up}$ and $\ket{\down}$ in the $m_J=0$, subspace.
(b) $^{87}$Sr  Zeeman diagram for the $2I+1 =10$ sublevels that asymptote to the $m_J=0$ subspace in the Paschen-Back regime (other subspaces not shown). Because of the large quadrupole effect, pairs with $\pm m_I$ are closely spaced for magnetic fields between 50 and 120 mT.  Qubits can be encoded in these pairs without loosing coherence in spontaneous emission.}
\label{fig:breitrabi}
\end{center}
\end{figure}

In order to obtain the product states of nuclear-electron spin degrees of freedom in $^1P_1$ as required for preserving nuclear spin coherence, we employ a sufficiently strong magnetic field.  The $^1P_1$ subspace is governed by the Hamiltonian including Zeeman interaction, magnetic spin coupling and quadrupole effects,
\begin{multline}
\hat{H}=A\hat{\v{I}}\cdot \hat{\v{J}}+Q \frac{3(\hat{\v{I}}\cdot \hat{\v{J}})^2+3/2 \hat{\v{I}}\cdot \hat{\v{J}}-I(I+1)J(J+1)}{2IJ(2I-1)(2J-1)}  \\
+g_J\mu_B \hat{\v{J}}\cdot \v{B}-g_I\mu_N \hat{\v{I}}\cdot \v{B},
\label{Hamiltonian}
\end{multline}
where $g_J$ and $g_I$ are the relevant electron and nuclear $g$-factors. The magnetic hyperfine constants are $A/h=-216$ Mhz and  $A/h=-3.4$ MHz for $^{171}$Yb and $^{87}$Sr respectively \cite{Boyd,Berends}.  For $^{171}$Yb, because $I=1/2$, the quadrupole constant $Q=0$ and we can analytically solve for the energy spectrum and eigenstates using a modified Breit-Rabi formula,
\begin{multline}
E_{m_J,m_I}=-\frac{E_{\text{HF}}}{2(2J+1)}+g_J\mu_B Bm_F \\
\pm\frac{E_{\text{HF}}}{2}\sqrt{1-\frac{4 m_F x}{2 J+1}+x^2},
\end{multline}
where $x=(g_I\mu_B+g_J\mu_N)B/E_{\text{HF}}$, $m_F=m_I+m_J$ and $E_{\text{HF}}=A(J+1/2)$. The resulting Zeeman diagram can be seen in Fig. \ref{fig:breitrabi}.  
The eigenstates are specified by $\ket{m_F}=\sum_q c_q(m_F) \ket{m_I=m_F-q}\ket{m_J=q}$, $q=0,1,-1$, with field dependent expansion coefficients $c_q$.  For high magnetic fields, in the subspace of interest, $c_0 \rightarrow 1$, and the states are in a product state $\ket{m_I}\otimes \ket{m_J=0}$. Deviations from this limit lead to the residual differential $g$-factor.

Since $^{87}$Sr has a large quadrupole constant $Q/h=39$ MHz \cite{Sauter74}, the Breit-Rabi formula does not apply.  Diagonalizing Eq.\ (\ref{Hamiltonian}) numerically gives the Zeeman diagram shown in Fig.\ \ref{fig:breitrabi} (for an analytic form, see \cite{Boyd}). Each of the three subspaces for $m_J=1,0,-1$ consists of 10 sublevels which asymptote to the $2I+1$ projections associated with the nuclear spin of this isotope, $I=9/2$.  For fields of order 10 mT or greater the Zeeman effect dominates over the hyperfine coupling and the sublevels approach product states of electron and nuclear spin.  However, the residual quadrupole interaction leads to a complex spectrum that is not described by a linear Zeeman shift with an effective $g$-factor.  For B-fields between 50 and 120 mT, the subspace with $m_J$=0 is nearly flat. Additionally, due to the quadrupole symmetry, the states with equal $|m_I|$ are paired and very close in energy (less than 2 MHz), whereas the energy splitting between the pairs is on the order of tens of megahertz, nearly as big as the linewidth.  Given the near degeneracy of the sublevels $\pm m_I$, we will consider preserving nuclear spin coherence in a qubit encoded in one of these two-dimensional subspaces, thereby ensuring that the frequencies of the different decay channels are not resolvable.

\begin{figure}
\begin{center}
\includegraphics[width=6cm]{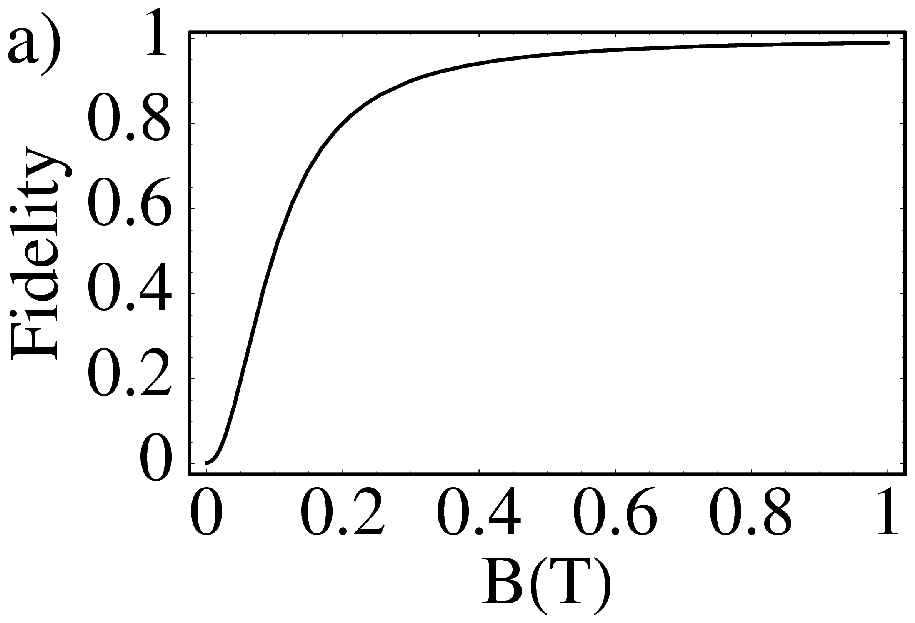}
\includegraphics[width=6cm]{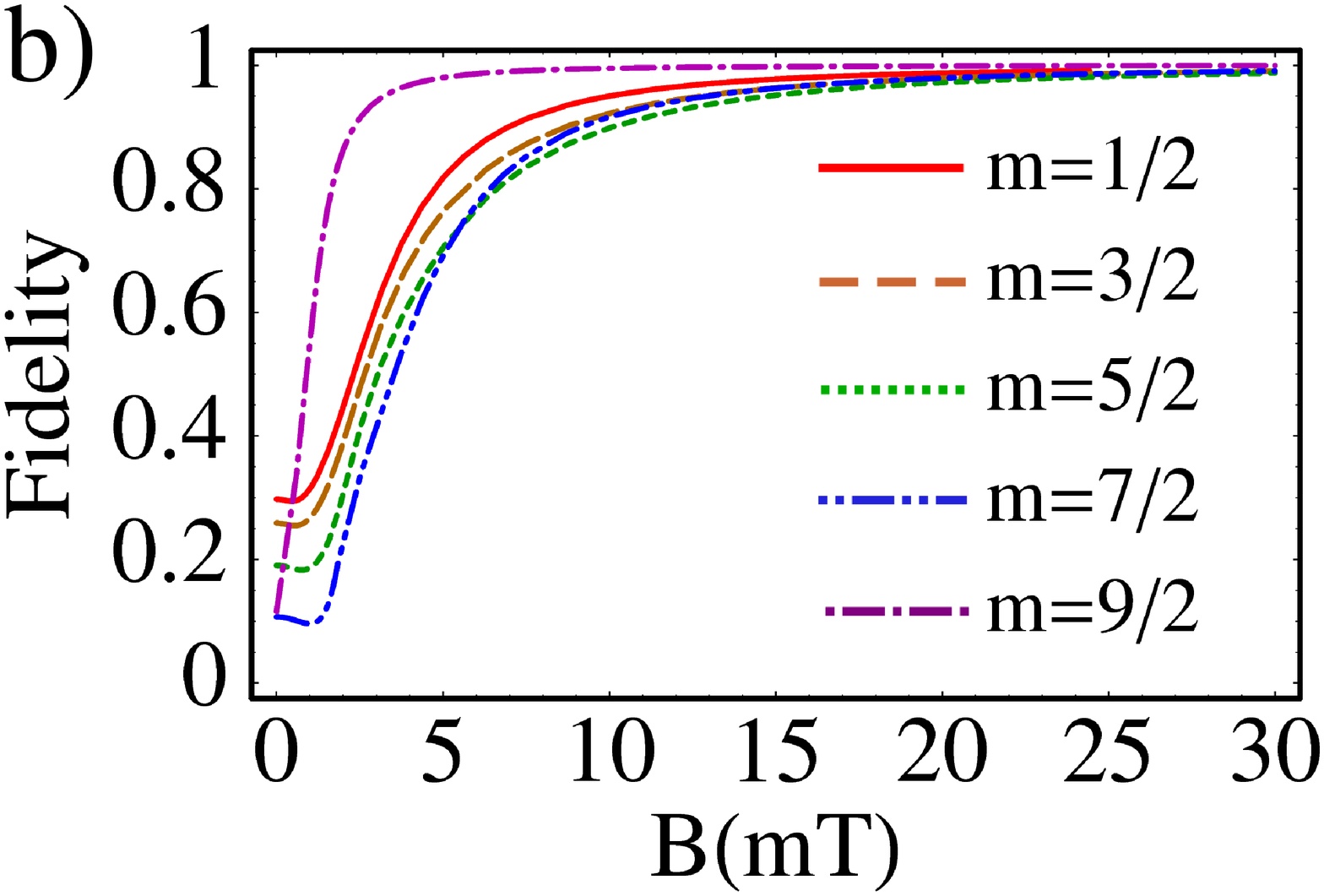}
\caption{(Color online) Fidelity for transfer of coherence, ${\cal F}=\Gamma'^2/(\Gamma^2+\delta^2)$ (see text), as a function of magnetic field. (a) The spin-1/2 qubit of $^{171}$Yb. (b) Different choices of qubit encoding as $\pm m_I$ in $^{87}$Sr.}
\label{fig:coherencetransfer}
\end{center}
\end{figure}

In the following we quantitatively study the transfer of nuclear spin-coherence via spontaneous emission. Consider a qubit in the excited state $\alpha\ket{e,\up}+\beta\ket{e,\down}$ which decays to the ground state $\ket{g}$.  We wish to transfer the qubit into the superposition $\alpha\ket{g,\up}+\beta\ket{g,\down}$. Here $\ket{e,\up(\down)}=\ket{^1P_1,m_F=\pm m_I}$ and  $\ket{g,\up(\down)}=\ket{^1S_0, \pm m_I}$.
The evolution of the atom can be described by a master equation in Lindblad form,
\begin{equation}
\dot{\rho}=-\frac{i}{\hbar}[H,\rho] -\frac{1}{2} \sum_q (L_q^\dagger L_q \rho +\rho L_q^\dagger L_q -2 L_q \rho L_q^\dagger ),
\end{equation}
where  $L_q= \sqrt{\Gamma} \sum_{m_F} c_q(m_F) \ket{g,m_I=m_F-q}\bra{e,m_F}$ (with $c_q$ defined above) are the ``jump operators" for the spontaneous emission of a photon with polarization $\pi, \sigma^+, \sigma^-$, $(q=0,+1,-1)$. Spin coherences, described by off-diagonal matrix elements, satisfy 
\begin{subequations}
\begin{align}
\dot{\rho}^{(e)}_{\up,\down} &= (-i\Delta_e - \Gamma) \rho^{(e)}_{\up,\down}\\
\dot{\rho}^{(g)}_{\up,\down} &= -i\Delta_g \rho^{(g)}_{\up,\down}+ \Gamma' \rho^{(e)}_{\up,\down}
\end{align}
\end{subequations}
where $\Delta_{e(g)}$ are the Zeeman splittings of the excited and ground qubits and $\Gamma' = \Gamma c_0(m_F) c_0(-m_F)$.  Solving for the ground state coherences in the limit $t\gg\Gamma$, when all population and coherence resides in the ground states,
\begin{equation}
{\rho}^{(g)}_{\up,\down}(t)= \rho^{(e)}_{\up,\down}(0) \frac{\Gamma'}{\Gamma - i \delta}e^{i\Delta_g t},
\end{equation}
where  $\delta=\Delta_g-\Delta_e$ is the differential energy shift.  The parameters $\delta$ and $\Gamma'-\Gamma$ are functions of applied magnetic field, both approaching zero in the perfect Paschen-Back limit. For finite fields, imperfect decoupling of electron and nuclear spin results in imperfect transfer of coherences, characterized by the ``fidelity" ${\cal F}=|\rho^{(g)}_{\up,\down}|^2/|\rho^{(e)}_{\up,\down}(0)|^2=\Gamma'^2/(\Gamma^2+\delta^2)$.  The resulting transfer of coherence as a function of the magnetic field can be seen in Fig.\ \ref{fig:coherencetransfer}. For $^{171}$Yb, a magnetic field of  1 T is required to reach a fidelity of 99\%, whereas for the $\pm m_I$ qubit in $^{87}$Sr a magnetics field of order 10 mT is sufficient to obtain this threshold.

Although we have argued throughout this Letter that one needs to decouple nuclear and electronic degrees of freedom in order to preserve nuclear spin coherence during laser cooling, we still require an interaction that allows for final readout of the nuclear spin state via resonance fluorescence. Consider first $^{171}$Yb.    According to Eq.\ (\ref{Hamiltonian}), the splitting between the $\ket{m_J,m_I=\pm1/2}$ states is given by $\Delta E=g_I\mu_NB+Am_J$, where $A$ is the hyperfine constant for the $^1P_1$ state.   Given $^{171}$Yb in a magnetic field of 1 T, the splitting between the $\ket{m_J\neq0,m_I=\pm1/2}$ states is on the order of 200 MHz. The splitting between the $\ket{\up}$ and $\ket{\down}$  in the $^1S_0$ ground state arises solely due to the Zeeman interaction with the nuclear spin, equal to 7 MHz for the applied field ($g_I \approx 1$).  This makes it possible to selectively drive the transition $\ket{^1S_0,m_I=+1/2}\rightarrow\ket{^1P_1,m_J=-1, m_I=+1/2}$ as shown in Fig.\ \ref{fig:overview} for readout.

For $^{87}$Sr in a magnetic field of few millitesla, the splitting between neighboring substates in $m_J = \pm 1$, dominated by the hyperfine interaction, is on the order of the linewidth and therefore not well resolved. Thus we cannot selectively excite a given $m_I$ level in the same way as we described for  $^{171}$Yb. We can, however, take advantage of the difference in the g-factors of the $^1S_0$ and $^3P_0$ states in Sr in order to manipulate population in the individual magnetic sublevels.  Consider a robust control pulse that transfers all population from $^1S_0$ to  $^3P_0$ level. The difference in Zeeman splitting implies that a narrow-band $\pi$-pulse can selectively return population in a chosen $m_I$ level to the ground state, leaving the remaining sublevels shelved in the metastable clock state.  The occupation of level $m_I$ can then be probed via fluorescence on the $^1S_0 \rightarrow {}^1P_1$ transition.  The procedure of shelving and activating the state of interest $m_I$, thus allows us to sequentially measure population in each sublevel.

We thank Chad Hoyt, Leo Hollberg, and Jun Ye for helpful discussions. This work was partly supported by Grants No.\ ARDA~DAAD19-01-1-0648 and No.\ ONR~N00014-03-1-0508

\end{document}